\documentstyle[preprint,aps,eqsecnum]{revtex}
\tighten
\begin{document}
\preprint{PITT-06-94}
\draft
%\pagestyle{myheadings}
%\markright{KINK  PRE ~~~~~~~\today  }

\title{Statics and Dynamics of an Interface\\
in a Temperature Gradient}
\author{$\mbox{D. Boyanovsky}^a, \mbox{David Jasnow}^a,
\mbox{J. Llamb\'{\i}as}^a \mbox{ and
F. Takakura}^{a,b}$}
\address{(a) Department of Physics and Astronomy\\University of
Pittsburgh, Pittsburgh, P.A. 15260, U.S.A\\
(b) Universidade Federal de Juiz de Fora, ICE,\\
Depto. de Fisica, Juiz de Fora, MG, Brazil
}
\date{}
\maketitle

\begin{abstract}
The response and nonconserved dynamics of a two-phase interface
in the presence of a temperature gradient oriented normally to the
interface are considered. Two types of boundary conditions
on the order parameter are considered,
and the structure of the effective free energy and
the Langevin equation for the
collective coordinate specifying the interface position are analyzed.
\end{abstract}


\section{Introduction}

Thermocapillarity effects, in which the interfacial properties
of two phase systems provide the dominant
driving forces and control
flows and motion, are of interest in a variety of areas.\cite{levich69}
These include two-phase flows, droplet
migration and a variety of other phenomena in a microgravity
environment (see, e.g., Refs.~\cite{hsu86,chhabra92} ),
and, for example, convection in which the Marangoni effect plays a
determining role (see, e.g., Ref.~\cite{thess94} and references cited).
Similar considerations could apply in
transport phenomena mediated by solitons
in systems with nonuniform temperature.
Thus our study may be of interest in charge-density wave
systems~\cite{gruner} and quasi-one dimensional organic polymers~\cite{YLU}
and other condensed matter systems in which solitons play an important
role.~\cite{bishop}
%~\cite{solitons}

To gain a deeper microscopic understanding into these phenomena,
it is of interest to consider the effect of a temperature gradient
on a two phase interface within a coarse-grained description.
We approach this question from both analytic and numerical
directions. The analysis presented here essentially
addresses the nature of the free energy of a two-phase
interface (or kink) in a temperature gradient. Accordingly,
the dynamics addressed here are restricted to a non-conserved order
parameter.

We find interesting features in the behavior. In particular,
a temperature gradient couples to the the so-called
``translation'' mode of the interface which is then
only weakly clamped by the finite size of the system.
A perturbative analysis has a vanishing domain of applicability
in a large system. The
structure of the effective free energy as a function of the
collective coordinate describing the interface position
is sensitive to the boundary conditions, but kinks not too
near the boundaries are seen to move with constant velocity
linear in the gradient.

The remainder of this paper is organized as follows.
In Section~\ref{sec:statics} the statics of a non-conserved order
parameter in a temperature gradient are considered, while in
Section~\ref{sec:dynamics} effective Langevin dynamics for the
collective coordinate
describing the interface position
are derived. A comparison
with a full numerical analysis using relaxational
(Model A in the lexicon of Ref.~\cite{HH})
dynamics is included.  Section~\ref{sec:concl} is reserved for
concluding remarks.

\section{Statics}
\label{sec:statics}

We consider a coarse-grained description of system allowing a
transition from a  single phase to a state of
two-phase coexistence with a scalar order
parameter, $\phi$.
%non-conserved order parameter $\phi$ (model A)
The temperature is allowed to vary slowly along one particular
direction (chosen to
be $x$) inside the sample,
which is confined in a box $0 \leq x \leq L$.
We consider configurations translationally invariant in directions
perpendicular to the gradient. For a configuration $\phi(x)$
the free
energy per unit `area' is taken to be
\begin{equation}
F= \int_{0}^{L} dx \left[ \frac{1}{2}
\left(\frac{\partial \phi(x)}{\partial x}\right)^2
-\frac{1}{2}r^2(x)\phi^2(x)+\frac{\lambda}{4}\phi^4(x)\right]\label{freenergy}
\end{equation}
In such a model
the function $r^2(x)$
can be taken to specify the local temperature
difference from a reference (critical) temperature.  We take
the temperature to
be slowly varying and parameterize
\begin{equation}
r(x)= r_0 \left(1+\frac{\Delta r}{r_0 L}x \right) \label{rofx}
\end{equation}
This parametrization will prove to be convenient for the analysis to follow.
Furthermore, for  $\Delta r /r_0 \ll 1$, which will be the case of interest,
the temperature varies almost linearly
with distance, with $x=0$ being the hotter and $x=L$ the colder ends of
the sample. While our explicit calculations and simulations are for
the specific model described above, general features of the results
are not expected to depend on the details of the free-energy
functional.

Equilibrium configurations are extrema of the free energy
functional (\ref{freenergy}),
which leads to the search for
solutions to the following non-linear differential equation:
\begin{equation}
-\frac{\partial^2\phi(x)}{\partial x^2}-r^2(x)\phi(x)+\lambda \phi^3(x)=0
\label{diffeq}
\end{equation}
Notice that with the particular choice (\ref{rofx}), $\phi(x)=\pm r(x)/
\sqrt{\lambda}$ is
an exact solution of (\ref{diffeq}) corresponding to a particular local
equilibrium configuration.
It proves convenient to
remove the local equilibrium variation and
to introduce a non-linear change of variables
\begin{equation}
\phi(x) = \frac{r(x)}{\sqrt{\lambda}}\eta(z(x))\label{eta}
\end{equation}
and the dimensionless parameter
\begin{equation}
h= \frac{\Delta r}{L r^2_0} \label{h}
\end{equation}
with $h \ll 1$.
(Variable changes of this type are discussed in Ref.~\cite{ince}.)
This parameter measures the strength of
the temperature gradient.
The dependence of $z(x)$ is determined by requiring
that the coefficient of the
$d^2\eta/dz^2$
in the resulting
differential equation for $\eta(z)$ be unity. This requirement
yields the simple relation
\begin{equation}
\frac{d z(x)}{dx} = r(x).
\end{equation}
We furthermore impose the boundary condition $z(0)=0$ thus obtaining the
new dimensionless variable
\begin{equation}
z(x)=r_0 x + \frac{h}{2}(r_0 x)^2
\end{equation}
measuring the `distance' from the hot wall.
With the assumption that $\Delta r / r_0 \ll 1$ and hence,
$ h , h Lr_0 \ll 1$, the
differential equation for
$\eta(z)$ in terms of the new dimensionless variable
$z$ becomes
\begin{equation}
\ddot{\eta}+3h\dot{\eta}+\eta-\eta^3=0 \label{etadiffeq}
\end{equation}
where dots stand for derivatives with respect to $z$. For $h=0$ there
are well known ``kink'' solutions to this equation (see e.g.,
Ref.~\cite{kinkrefs}).

The advantage of parametrizing $r(x)$ as in (\ref{rofx}) and of the change
of variables (\ref{eta}) becomes clear. Whereas in the original
differential equation
(\ref{diffeq}) the $x$ dependence of $r(x)$ broke explicit translational
invariance, the differential equation for $\eta(z)$ is {\it manifestly}
translational invariant in the variable $z$. Furthermore,
Eq.~(\ref{etadiffeq})
provides a very appealing physical interpretation: $\eta(z)$ describes the
trajectory of a particle moving in ``time'' $z$ in a potential
\begin{equation}
V(\eta) = \frac{1}{2}\eta^2-\frac{1}{4}\eta^4 \label{potential}
\end{equation}
damped by constant friction proportional to $h$. The present
work involves analysis of Eq.~(\ref{etadiffeq}).

Before searching for solutions to the differential equation, we
must specify boundary conditions. Since we are interested in the
behavior of interfaces, relevant boundary conditions are
those compatible with solutions that have one node (the position of the
interface).
We will focus on two sets of boundary conditions.
The first, Type I, corresponds to natural boundary
conditions:
\begin{equation}
\mbox{Type I:} \; \; \;
\frac{d\phi(x)}{dx}\mid_{x=0} = 0 \; \; ; \; \;
\frac{d\phi(x)}{dx}\mid_{x=L} = 0. \label{deribc}
\end{equation}
The corresponding conditions on $\eta(z)$ at $z=0 \mbox{ and }
z = l \equiv z(x=L)$
are $\dot{\eta}(z=0,l) = -h \eta(z=0,l)$.
Note we always will consider $hLr_0 << 1 $.

In the boundary conditions of Type II, the order parameter is fixed
at the ``local equilibrium'' values, namely
%II:)
\begin{equation}
\mbox{Type II:} \; \; \;
\phi(0) = \mp \frac{r(0)}{\sqrt{\lambda}} \; \; ; \; \;
\phi(L) = \pm \frac{r(L)}{\sqrt{\lambda}} \label{bc}
\end{equation}
corresponding to $\eta(0) = \mp 1 \; \; ; \; \; \eta(l) = \pm 1.$

\subsection{Boundary Conditions I}
\subsubsection{Perturbative solutions}

For these boundary conditions there is surely a solution in which
the order parameter remains in one phase, with the
$|\eta| \simeq 1$. We refer to this as the
local equilibrium solution, which will be a global minimum of the
free energy functional. For small $h$
a perturbative solution is easily found with $\eta(z) = -1+ \delta(z)$
and
$\delta(z) =  e^{-\frac{3}{2}hz}\left( A \cosh(Wz)+ B \sinh(Wz)\right)$,
with $A,B$ chosen to satisfy the boundary conditions and
$W = \sqrt{2+\frac{9}{4}h^2}$.

Having established the configuration with the lowest free energy for
the case of boundary conditions (I), we now concentrate on finding
a solution with an interface. This solution must necessarily have one
node, and for very small $h$, we expect that its  behavior
away from the node will have  $|\eta| \simeq 1$.

For $h=0$, the solution $\eta_0(z)$
may be found by quadratures in terms of the
elliptic sine function\cite{byrd}.
For $h \neq 0$ there are no exact solutions to the differential equation,
which in its original form (\ref{diffeq}) may be recognized
(for $h \ll 1$) as a Painlev\'{e} transcendental without an
exact solution\cite{ince}. Although this is a formal property,
one of its main  consequences is that the solution is non-singular in a
finite interval\cite{ince} (the only possible singularities are at
infinity), and
this suggests that a perturbative approach may be feasible, namely
letting
\begin{equation}
\eta(z) = \eta_0(z)+h\eta_1(z)+ \cdots \label{pert}
\end{equation}
This expansion will allow us to study the linear response of the interface
profile to the temperature gradient.

We now need to find the first order correction $\eta_1$. This function obeys
the linearized equation
\begin{equation}
\ddot{\eta}_1+\eta_1-3\eta_0^2 \eta_1= -3h\dot{\eta}_0 \label{lineareq}
\end{equation}
This shows that the temperature gradient couples to the ``translation
mode'' of the system, $\delta_1 \equiv \dot{\eta_0}$, which satisfies
the homogeneous equation in Eq.~(\ref{lineareq}). The existence of
this
mode yielding zero eigenvalue for the fluctuation operator
(${\cal L} = d^2/dz^2 + 1 - 3 \eta_0^2 $)
is a direct result of translational invariance (see e.g., Ref.~\cite{jasnow}).
Were it not for the fact that the domain is finite, such a
perturbative analysis could not be carried out.
In the present case a solution can be constructed
using elementary methods.
{}From the solution $\delta_1$
one can construct another linearly independent solution of
the homogeneous equation, $\delta_2$,
with unit Wronskian. Some additional steps are included in Appendix A.
%\delta_1(Z) & = & \dot{\eta}_0(Z) \label{sol1} \\
%\delta_2(Z) & = & \delta_1(Z) \int^Z_{0} \frac{dZ'}{\delta^2_1(Z')}
%\label{sol2}
%\end{eqnarray}
As one might expect, near $Z=z-l/2 \simeq 0$ the solution is of the
form
\begin{equation}
\eta(Z)=\eta_0(Z)+h \alpha_1 \frac{d \eta_0(Z)}{dZ} + \cdots \approx
\eta_0(Z+h\alpha_1)+ \cdots \; , \label{trans}
\end{equation}
where the constant $\alpha_1$
(see Appendix A)
is determined by the boundary conditions,
and the terms indicated by dots vanish at $Z=0$.
Clearly the position of the interface has been shifted to $Z=-h \alpha_1$
(i.e., $z=(l/2)-h \alpha_1$); it remains to calculate $\alpha_1.$
A perturbative analysis will be possible as long as $h \alpha_1$ is
sufficiently small.

The explicit evaluation of $\eta_1$ is, in general,
complicated by the unwieldly elliptic functions and
elliptic integrals. However, it is simplified in the physically relevant
large volume limit.
Some background is provided in Appendix B.
Since we want only one node of the order parameter profile, this requires
that the half-period of the elliptic function that
corresponds to the unperturbed solution (\ref{elliptic}) becomes very
large in the large volume limit.
After some tedious but straightforward
algebra we find that
\begin{equation}
\alpha_1 \approx -\frac{\sqrt{2}}{64} e^{\sqrt{2}l}. \label{alpha1bcI}
\end{equation}

Before proceeding it is useful to consider the perturbative
correction in the case of small amplitudes. That is, one may
linearize Eq.~(\ref{etadiffeq}) near
$\eta \approx 0$, discarding the cubic term.  Although in this approximation
the differential equation describes an (underdamped) harmonic oscillator,
the linear response field $\eta_1(Z)$ can be  compared to
the exact solution. Requiring that there be only one node in the
interval requires that  $l=\pi$ (the
asymptotic limit of the elliptic integral $K(m)$; see Appendix B).
The first order correction $\eta_1$  agrees with the first order term
(in $h$)
in the expansion of the exact solution with boundary conditions
of Type I.
The amplitude of the perturbation
is of order $hl$,
and the perturbative
analysis is reliable whenever $ h l \ll 1$, corresponding physically to the
gradient being small on the scale of the correlation length.
Ultimately secular terms destroy the approximation as $ h l $ becomes
too large. The situation is much worse in the non-linear case
to which we now return.

The result contained in Eq.~(\ref{alpha1bcI})
is important: the coefficient that determines the translation
of the interface becomes exponentially large in the large volume limit,
signaling the breakdown of perturbation theory. The relevant combination
can be seen from Eq.~(\ref{trans}) to be $h \alpha_1$.
Furthermore,
the sign of the coefficient
shows that the interface is shifted dramatically towards the colder end.
This suggests
that the extremum found does not correspond to a local minimum of
the free energy functional.
However, the result is physically reasonable. In the very large
volume limit the translation mode is only weakly clamped by the boundaries,
and since, as we
argued above, the temperature gradient couples to this mode,
the result is
a large shift of the interface under the perturbation.

\subsubsection{Numerical results}

Since exact analytical solutions are not generally available
and perturbation theory
leads to an exponentially divergent linear response to the temperature
gradient, we obtained numerically the profiles for the order parameter using
a shooting method to solve the differential equation with
Type I (natural) boundary conditions given in Eq.~(\ref{deribc}).

Figure~1 shows the $\eta(z) \mbox{ vs } z$ for $h=10^{-3}$ for a system
of size $l\approx 40$. We clearly see that the perturbed interface
is established near
the cold end of the sample, forming a boundary layer of about two
correlation lengths. (In dimensionless units, the correlation length is
$\xi \approx \sqrt{2}$). From the numerical standpoint
and the interpretation
of $\eta(z)$ as the trajectory of a particle rolling down the potential
(\ref{potential}) under constant friction $h \ll 1 $, the reason that the
extremum configuration has an interface
close to the cold end in the large volume limit $l \gg \xi$ is clear.

The initial condition for ``shooting'', that is the value of $\eta(0)$ for
the boundary condition $\dot{\eta}(z)|_{z=0} = -h \eta(0)$, must be such
that $\eta(0) < -1 \; ; \; \dot{\eta}(0) > 0$. In this manner, the
particle first has to climb up
the potential hill to $\eta = -1$ and reach that point with an extremely
small velocity, remaining for a long time around that point and slowly
falling
down on the other side of the potential maximum.
It then increases its velocity,
passes through $\eta =0$ and climbs up the potential hill
towards $\eta = +1$.
Because of friction, the velocity reaches zero before reaching the top, and
the particle turns back; the integration stops when the boundary condition
is obeyed again, now with negative velocity.
On the other hand,
if the integration were begun with $\eta(0) > -1$,
for a large volume there would only be solutions with many nodes
corresponding to a higher free energy.
A solution with only one node  and $\eta(0) > -1$ will only
appear for small volumes, compatible with the solution found in the
linearized region near $\eta \approx 0$.

Thus in the large volume limit, the particle must begin with $\eta(0) < -1$
(the ``antikink'' solution will begin with $\eta(0) >1$) with a small
upward velocity given by the boundary condition and will remain near the
local equilibrium region $\eta \approx -1$ for most of the time, making
a rather quick transition near the cold end $z \approx l$.
However unlike in the topological ``kink'' case, the solution with one
interface
is in the same sector in functional space  as the local
equilibrium solution in that it has the
same boundary conditions
(of Type I). But clearly the local equilibrium solution
corresponds
to the lowest free energy amongst the functions with
such boundary conditions.
The solutions to the differential equations (\ref{diffeq})
and (\ref{etadiffeq})
are indeed
extrema of the free energy functional; the solution with one interface
cannot be a
local
maximum because  adding  ``wiggles'' in the
configuration will
increase the free energy via the derivative terms.

If the configuration were a {\it local} minimum with a free energy
higher than
the local equilibrium solution, then there must be a local maximum that
separates the two solutions.
However there is no evidence of another solution
with the same boundary conditions. This reasoning leads us to conjecture
that
the solution with the interface is most likely a saddle point of the free
energy functional. To prove this conjecture, we would have to study the
spectrum of fluctuations around this solution with an interface and identify
a particular direction in function space for which an eigenvalue is
negative.
In the present case this is an extremely difficult problem complicated by
the boundary conditions on the solution.

In the case of zero temperature gradient and in the large volume limit,
because of translational invariance, a
shift of the interface costs negligible free energy (when the
interface is far from the boundaries).  With the temperature gradient,
translational invariance is
broken and there is a profile which
extremizes the free energy. Thus our strategy is to propose a good trial
``kink-like'' function parametrized by the position of the interface $z_c$
and to compute the free energy as a function of this parameter. This
 is equivalent to treating the position of the interface
 as a ``collective coordinate,'' which is
appropriate in
the case of kinks and identifies the coordinate, $z_c$,
as the translational
degree of freedom\cite{jasnow,jasrud,sakita}.

We have found  for a wide range of
parameters
$h \mbox{ and } l$ that the  numerical solution
to the differential equation (\ref{etadiffeq}) is very well described
by the
interpolating function
\begin{equation}
\eta(z,z_c) \approx (1+\frac{h}{\sqrt{2}})
\tanh\left[\frac{z-z_c}{\sqrt{2}}\right].
\label{kink}
\end{equation}
Here $z_c$ determines the position of the interface. The accuracy of this
fitting function is better than $1\%$ in most of the volume, with slightly
larger departures of about $2-3\%$ near the boundaries of the sample, but
extremely accurate near the interface.
In terms of  $\eta(z;z_c)$ and the variable $z$, the free energy as a
function of $z_c$ is given  by [up to linear order in $h$ consistently
with our
expansion of the differential equation (\ref{etadiffeq})]
\begin{eqnarray}
F[z_c] & & \approx \left(\frac{r_0^3}{\lambda}\right)
\int_0^l dz (1+3hz)\left\{\frac{1}{2}
\left(\frac{d\eta(z;z_c)}{dz}\right)^2 - \frac{1}{2}\eta^2(z;z_c)+\frac{1}{4}
\eta^4(z;z_c)\right\} \nonumber \\
      & & +\frac{h}{2}\left(\eta^2(l;z_c)-\eta^2(0;z_c)\right)
\label{freenergyzc}
\end{eqnarray}
It is clear from this expression that in terms of the field $\eta$ and the
variable $z$, the breakdown of translational invariance is in the metric
(and for large enough system, weakly from the boundaries).

Figure 2 shows $F[z_c] \mbox{ vs } z_c$ for
$h=10^{-3} \; , \; l \approx 40$
obtained by using (\ref{kink}); these values of the parameters are the same
as those for Figure~1 (Type I boundary conditions).
We see that $F[z_c]$ has a {\it maximum} at $z_c = z_{max} \approx 36$
whereas the
``shooting''
numerical integration gives the value of the position of the interface
(i.e., the node, $\eta = 0$) at $z_c = 35.8$, giving confidence
that the full numerical approach yields a solution corresponding
to this maximum.
Except within a few correlation lengths of the boundaries,
we find that $F[z_c]$ varies approximately
linearly with the interface position $z_c$ with a slope
\begin{equation}
\frac{d F[z_c]}{dz_c} = \left(\frac{r_0^3}{\lambda}\right)2\sqrt{2}h +
{\cal{O}}(h^2)+\cdots
= 3h \sigma_0 +  {\cal{O}}(h^2)+\cdots \label{slope}
\end{equation}
where $\sigma_0 = r_0^3 2\sqrt{2}/3 \lambda$ is
recognized as the interfacial free energy (surface tension) for an
interface when $h=0$.
We will compare this to our dynamical simulations described below.

These and other numerical consistency checks between the trial
function approach and the full nonlinear solution leads us to conclude
that the configuration that extremizes the free energy with one
interface (node)
corresponds to a {\it maximum} in the
functional
direction corresponding to translations, and is thus interpreted as a
saddle point configuration. This saddle already exists for the
unperturbed case
($h = 0$) under the boundary conditions of Type I, and the situation
for small $h$ represents a smooth deformation.

If this interpretation is correct, there emerges the question
as to the identification of
the thermodynamically different states separated by this saddle.
These states should
be global minima of the free energy functional, because if one were a
global and
the other a local minimum, there should then be
%at least {\it three}
additional solutions
of the differential equation with different free energies. However,
 as mentioned above, we  find only two: the local equilibrium solutions
(near $\eta = \pm 1$) and the interface solution.
Thus our conclusion is that the saddle point separates the
thermodynamically
different states corresponding to the
(nodeless) local equilibrium solutions
near $\eta = \pm 1$, which are degenerate.
This interpretation will be
strengthened by the study of the dynamics in the next section.

\subsection{Boundary conditions of Type II}
\subsubsection{Perturbative analysis}

These are ``topological'' boundary conditions that force the
order parameter
to have at least one node, and clearly there is no equivalent of the
local equilibrium configuration (in which
the order parameter maintains the same sign) that is available with
boundary conditions I.
Hence we expect that
the single-node solution of the differential equation (\ref{etadiffeq})
with
boundary conditions of Type II is thus an absolute minimum of the
free energy in
the space of functions with these boundary conditions.
Before analyzing
the solution numerically,
it proves illuminating to study the linear response as in the
previous case. For sufficiently large volumes an excellent
approximation to the
unperturbed solution ($h=0$) is $\eta_0(Z)=\tanh[Z/\sqrt{2}]$. With this
unperturbed solution we can explicitly construct the functions
$\delta_1(Z) \; ; \; \delta_2(Z)$ and the response field $\eta_1$
satisfying $\eta_1(-l/2)=\eta_1(l/2)=0$
(see Appendix A).
Again because the boundary conditions are symmetric,
the coefficient $\alpha_2$ vanishes, and we find (for $ l \gg 1$)
\begin{equation}
\alpha_1 \approx \frac{e^{\sqrt{2}l}}{16}. \label{alfa1bcII}
\end{equation}
Perturbation theory has a vanishing domain,
but an important feature is  that, in contrast with the previous case,
now $\alpha_1$ is {\it positive}. This means that in this case the
perturbed interface
lies closer to the hot end of the sample.

\subsubsection{Numerical results}

We have used the same numerical scheme to solve the
full nonlinear differential
equation (\ref{etadiffeq}), now
beginning with $\eta(0)=-1$ and ``shooting'' with an initial
derivative such that
$\eta(l)=1$. Figure~1 also shows the profile for $h=10^{-3} \; ; \; l=25$ (Type
II
boundary conditions).

The profile  is  again easily understood in terms
of the particle rolling down the potential hill in the presence
of friction: the particle has to begin from $\eta(0)=-1$ but with a fairly
large derivative because of the friction term. If the initial velocity is
small then the particle does not make it up the hill to reach $\eta=1$
% in time $l$ because of friction.
Thus the initial velocity is fairly large and
the particle moves very rapidly initially taking a short time to
reach $\eta=0$; hence the interface is very close to $z=0$,
the
hot end. Eventually the particle climbs up the potential hill,
being slowed
down not only by the potential but also by the friction.

One knows that, because of the coupling to the translation mode,
the linear response must necessarily diverge in the
infinite volume limit. Here, for a large but finite system, the
position of the perturbed interface is very close
(about 2-3 correlation lengths)
to the hot end, far from the unperturbed value at the middle
of the system.
The sign of the translation is correctly predicted by the linear response
calculation as is the case for boundary conditions of Type I.
However,
unlike the case of Type I boundary conditions, this solution, as discussed
above,
corresponds to the lowest free energy compatible
with the odd boundary conditions of Type II.

\section{Dynamics}
\label{sec:dynamics}
\subsection{Langevin equation for the collective coordinate}

Below we will present the results of simulations of
relaxational dynamics
for the motion of a two phase interface in a temperature gradient.
Before doing so, we derive from the trial function (\ref{kink})
the velocity of the interface.

Relaxational dynamics for this non-conserved order parameter
are specified by assuming
a Langevin equation description
\begin{equation}
 \frac{\partial \phi(x,\tau)}{\partial \tau} =
 - \frac{\delta F}{\delta \phi}.
\label{langeq}
\end{equation}
For our present purposes we neglect a noise term and absorb the
characteristic relaxation rate into the dimensionless time, $\tau$.
Assuming that there is a very small distortion of the profile
as a function
of time, that is, that the time evolution corresponds to translations of
the interface, we propose the parametrization
\begin{equation}
\phi(x,\tau) = \frac{r(z)}{\sqrt{\lambda}}\eta(z-z_c(\tau))
\end{equation}
This parametrization leads at once to
an equation for
the ``collective coordinate'' $z_c(\tau)$.
For $\xi \ll z_c \ll l \mbox{ and } h \ll 1$ and using the
trial function (\ref{kink})
we find
\begin{equation}
\frac{d z_c}{d\tau} = -3r_0^2h + {\cal{O}}(h^2)+\cdots \label{velocity}
\end{equation}
The interface is predicted to move with constant speed proportional
to the gradient; in the language of solitons the coefficient would
be identified as the kink (linear) mobility.
Details are compared directly with the results of simulations below.

\subsection{Numerical simulation}
We follow the time evolution of the system using the Langevin dynamics,
(\ref{langeq}). These dynamics drive the system to a free energy minimum.
%\begin{equation}
Rescaling $\phi$, $x$ and $\tau$ to eliminate inconsequential
parameters, we obtain
\begin{equation}
\frac{\partial \phi(x,t)}{\partial t} =
\frac{\partial^2\phi(x,t)}{\partial x^2}+
(1+ h x)^2\phi(x,t)-\phi^3(x,t),
\label{eveq}
\end{equation}
where we use the same symbols for the rescaled variables for simplicity.
The only parameter remaining is $h$,
which corresponds, as above, to a temperature
gradient, such that temperature decreases with increasing $x$ if
$h$ is positive.

The equation (\ref{eveq}) was numerically integrated using a simple Euler
discretization on a one dimensional lattice of $100$ nodes with mesh
size $\Delta x= 0.1$ and time step $\Delta t=0.001$.
We considered separately the two types of boundary conditions discussed
above. In Type I, the
order parameter $\phi$ was required to have zero gradient at the
boundaries by imposing reflecting boundary conditions. In the second
case (Type II), the value of the order parameter was fixed
at the boundaries,
with values corresponding to a different phase at each end of the
sample.
The value chosen was the
equilibrium  order parameter for an isothermal system
at the temperature corresponding to that point.
(This corresponds to $\eta = \pm1$ in the earlier parameterization.)
The initial conditions used were of the form
\begin{equation}
\phi(x,0) = (1+h x)\tanh\left[\frac{x-x_0}{\sqrt{2}}\right] \, ,
\label{tanh}
\end{equation}
where $x_0$ is the initial position of
the kink. This function is very close to the actual values that the
order parameter takes once it enters the dynamic regime,
as long as  $x_0$ is not too close to the boundaries.

For the open boundary case (Type I), it was observed that the
evolution of the
system consists of the kink being displaced until it disappears at
one of the boundaries. The final state is then always a single phase
in the whole system, and at that point there is no further change in the
order parameter. This type of solution has been referred
to above as the `local equilibrium' configuration.
This is expected, since the one-phase configuration
constitutes the global minimum of the free energy, and it can be reached
with the imposed boundary conditions.

For most initial positions the kink moves towards the higher
temperature side.
This is to be expected since near the high temperature side
the correlation
length is larger and the equilibrium
order-parameter gap is smaller. These effects decrease
the interfacial free energy so that the kink can evolve to a
lower free energy
state by moving toward the hot end.
However, it is interesting that if the initial kink position is close
enough to the
cold boundary, the kink disappears at that end. Figures~3(a)
and 3(b) show
the evolution of the kink when it is started from two different
positions; in all cases the kink disappears at the boundary.
This can be understood by looking at the free energy of
the system as a function of the kink position $x_0(t)$, defined as
the point where the order parameter vanishes. The free energy has a
maximum at some value of $x_0$ close to the cold end.
The larger the temperature
gradient $h$, the closer  this maximum is to
the cold end.
Figure~4 shows this free energy as a function of the kink position. It
was calculated from equation (\ref{freenergy}) applied to the
the order parameter configuration as the kink evolves. For a
given initial condition, we can obtain only a portion of the graph
as the interface evolves toward one side.
However, for
different initial conditions, the calculated free energy always falls
on the same curve, as expected.

For the case of fixed boundary conditions, i.e., Type II,
the free energy has a
minimum at some point close to the higher temperature side, and
therefore the kink moves towards that point from any initial
position. The free energy for this type of boundary conditions is
also shown in Figure~4.

Finally, also shown in
Figure~4 is the free energy calculated for
configurations of the order parameter given by
the trial function (\ref{tanh}) for
varying $x_0$. As can be seen, the two curves obtained dynamically
with differing boundary conditions
coincide in the region away from the boundaries, which shows that the
dynamics of the kink will not be affected by the boundary conditions
until the kink comes within a few correlation lengths
of the boundaries. The curve obtained from the local
temperature solution (\ref{tanh}) is also very close to the others in
the region away from the boundaries.

To compare with the analytical results of equation (\ref{slope}), we
computed the value of $d F[x_0]/dx_0$ for small values of
$h$, in the region far from the boundaries, which is where the approximate
analytical results are expected to hold.
The numerical and analytical results for the slope of the
effective free energy of the interfacial configuration agree within
2\%.
As shown in (\ref{velocity}) the velocity of the interface is
expected to be proportional to the gradient. Our numerical simulations
yield the coefficient within 2-3\% of the approximate analytic
result. We conclude that as long as the system size is
sufficiently large (here ten correlation
lengths is seen to be large enough), the trial function provides a
semiquantitative
description of the statics and relaxational dynamics.

\section{Concluding remarks}
\label{sec:concl}
In this short paper we have considered the statics and non-conserved
relaxational dynamics of a two-phase interface in which the system is
subjected to a temperature gradient. For the statics one has to
extremize the free energy; a non-linear change of variables
preserves translational invariance and yields a representation
in which the interface shape becomes equivalent to
the time trajectory of a ball rolling in a potential, but slowed
by friction (Eq.~(\ref{etadiffeq})).

It is seen that the temperature gradient couples to the `translation'
mode of the unperturbed (isothermal) interface, so that perturbation theory
can be applied to the introduction of the temperature gradient only if the
mode is clamped  by the finite size of the system. Even then
the clamping is exceptionally weak
for large systems, and the `linear response' is
divergent, but correctly predicts which side of the sample
will contain the interface.
In realistic terms perturbation theory has vanishing domain since
the effective coupling becomes $ \sim h \exp(l)$,
where $h$ is a measure
of the gradient and  $l$ is  a measure of the system size,
in units of the correlation length. Hence full nonlinear
solutions and dynamical simulations were carried out numerically.

The equilibrium configurations
are sensitive to the boundary conditions. For Type I boundary
conditions in which the order parameter derivative vanishes at the
walls, the global equilibrium is for the system to remain in one
phase.
A single-kink extremum is argued to be a saddle point in the space
of functions satisfying these boundary conditions.
Simulations using relaxational (Langevin) dynamics
reveal that an initial interface travels to one
of the walls and the interface disappears. An interesting feature
is that, generally, the interface travels toward the hotter wall at
approximately constant velocity;
however, if the initial interface is established close enough to the
colder wall, it is removed there.

For `topological' boundary conditions (Type II) in which the
order parameter is
forced to be in two different phases at the ends of the system, the
equilibrium configuration has the interface near the hotter wall,
which is consistent with the behavior of the free energy.

The structure of the free energy in the space of single kink configurations
has been analyzed approximately using trial functions which should
be accurate as long as the system size is sufficiently larger than the
thermal correlation length. Good numerical agreement is found with
purely numerical relaxational dynamics. The trial function
along with assumed relaxational dynamics yields analytic results for
the equation of motion of the interface position (collective coordinate),
also in good agreement with purely numerical simualtions.

For an order parameter with conserved (say, Model B~\cite{HH}) dynamics
the situation is different. A configuration with a single interface cannot
change much in response to a temperature gradient. However, a
kink-antikink pair, representing a slab or bubble of one
phase in the other can move significantly while respecting the
conservation. This topic is beyond the present scope and will
be explored elsewhere.

\vspace{0.25in}

\noindent {\bf Acknowledgments}
D.B. thanks H. J. de Vega for illuminating discussions, and he
is grateful to the National Science Foundation for support
under grant PHY-93-02534 and INT-9016254 (Binational Collaboration
with Brazil). F.T. has been supported by CNPQ and thanks the
Department of  Physics and Astronomy
of the University of Pittsburgh for its hospitality.
D.J. and J.L. are grateful to the Microgravity Science and Applications
Division of NASA for support of the work under grant NAG3-1403.

\appendix
\section{ Formal solution of the linear response}
The perturbation $\eta_1$ satisfies Eq.~(\ref{lineareq}). Now,
$\eta_0$ is a function of $Z=z-l/2$, so $\eta_1$ is also a function of
this variable.
As noted in the text the function
$\delta_1(Z)=\dot{\eta}_0(Z)$ is
an eigenfunction of the second order fluctuation operator with eigenvalue
zero.
With this solution
we can construct another linearly independent
with unit Wronskian. Thus we find
the following  solutions of the homogeneous equation:
\begin{eqnarray}
\delta_1(Z) & = & \dot{\eta}_0(Z) \label{sol1} \nonumber \\
\delta_2(Z) & = & \delta_1(Z) \int^Z_{0} \frac{dZ'}{\delta^2_1(Z')}
\label{sol2}
\end{eqnarray}

The zero mode, $\delta_1(Z)$ is a symmetric function around
$Z=0$ and vanishes (linearly) at $Z=\pm l/2$.
This motivates the choice of the lower limit
in (\ref{sol2}), since now $\delta_2(Z)$ is antisymmetric
around
$Z=0$ and obviously finite at $Z=\pm l/2$.
Finally, the solution to Eq.~(\ref{lineareq})
is
\begin{equation}
\eta_1(Z) = 3\delta_1(Z)\int_{0}^Z dZ' \delta_1(Z')\delta_2(Z')-
3\delta_2(Z)\int_{0}^Z dZ' \delta^2_1(Z')+\alpha_1 \delta_1(Z) + \alpha_2
\delta_2(Z) \label{responsefield}
\end{equation}
with constants $\alpha_{1,2}$ to be determined by the boundary
conditions. Expanding the
set of Type I boundary conditions to linear order in
$h$ (and accounting for the explicit $h$-dependence of the variable Z),
one finds that the boundary conditions are symmetric
around $Z=0$, and since
$\delta_2(Z)$ is antisymmetric, the coefficient $\alpha_2$ vanishes.
(This
is a bonus of the parametrization in terms of Z and the choice of
lower limit
of the integrals in the functions above.)
This indicates that
the position of the interface is shifted;
analyzing the behavior near $Z \simeq 0$ yields Eq.~(\ref{trans}).

\section{Unperturbed kink}
In a finite box the solution of Eq.~(\ref{etadiffeq})
for $h=0$ may be found by quadratures in terms of the
 elliptic sine function\cite{byrd}
\begin{eqnarray}
\eta_0(z) & = & \eta_i {\rm sn}[u|m] \label{elliptic} \nonumber \\
u         & = & \left(z-\frac{l}{2}\right) \sqrt{\frac{2-\eta_i^2}{2}}
\nonumber \\
% \label{u} \\
m         & = & \frac{\eta_i^2}{2-\eta_i^2},
% \label{m}
\end{eqnarray}
where $\eta_0(0) = -\eta_i = -\eta_0(l)$ gives the values at the
end points of the interval.
The requirement that there is only one node in the interval is equivalent
to requiring that the half-period of this solution be identified with
$l$. This requirement in turn determines the value of $\eta_i$ from the
relation
\begin{equation}
K(m)=\frac{l}{2}\sqrt{\frac{2-\eta_i^2}{2}} \label{quarterperiod} \nonumber
\end{equation}
in which $K(m)$ is the elliptic integral of the first kind~\cite{byrd}
and is a quarter period of the elliptic function $sn$.
This solution
is a function of $Z=z-l/2$, and it obeys
$\dot{\eta}_0(Z)|_{Z=-l/2}=\dot{\eta}_0(Z)|_{Z=l/2}=0$.

%\end{document}

\newpage

\begin{center}
\underline{\bf Figure Captions}
\end{center}

\begin{enumerate}
\item[Figure 1]
Solid line: $\eta(z) \mbox{ vs } z \; \; ; \; \; h=10^{-3}\; \; ; \; \;
l\approx 40$ with boundary conditions (I), showing the
extremum solution with the interface near the cold boundary.
Dashed line: $\eta(z) \mbox{ vs } z \; \; ; \; \; h=10^{-3}\; \; ; \; \;
l\approx 25$ with boundary conditions (II), showing the
extremum (global equilibrium) solution with the interface near
the hot boundary.
\item[Figure 2]
 $F[z_c] \mbox{ vs } z_c$ for $h=10^{-3} \; ; \; l \approx 40$
with the parametrization (\ref{kink}).
\item[Figure 3]
(a) Dynamical evolution of an interfacial
structure showing the disappearance of the interface at the hot boundary.
(b) Evolution of structure with
initial interface near the cold boundary showing
the disappearance of the interface there.
\item[Figure 4]
Effective free energy as a function of the collective coordinate
(position of the node in the order parameter profile) showing,
for comparison, the free energy for the hyperbolic tangent trial
function. In all cases $h = 0.002.$
\end{enumerate}

\end{document}